\documentclass[12pt,preprint,dvips,iop]{emulateapj}
\usepackage[backref,breaklinks,colorlinks,citecolor=blue]{hyperref}
\def \msun{$\mathrm{M}_\odot$}%
\def \kms{km~s$^{-1}$}%
\def \pap1{Paper I}%
\def \pap2{Paper II}%
\def \deg{$^\circ$}%
\def \hb{heartbeat stars}%

\def\K#1{\ifnum#1=1{{KIC~3230227}}%
\else\ifnum#1=2{{KIC~3749404}}%
\else\ifnum#1=3{{KIC~4248941}}%
\else\ifnum#1=4{{KIC~8719324}}%
\else\ifnum#1=5{{KIC~9899216}}%
\else\ifnum#1=6{{KIC~11494130}}%
\else\relax \fi\fi\fi\fi\fi\fi}%

\shorttitle{Heartbeat Stars}
\shortauthors{Smullen \& Kobulnicky}
\slugcomment{{\sc Accepted to \apj:} 8 June 2015 } 

\begin{document}
\title{Heartbeat Stars: Spectroscopic Orbital Solutions for Six Eccentric Binary Systems}
\author{Rachel~A.~Smullen\altaffilmark{1,2}}
\author{Henry~A.~Kobulnicky\altaffilmark{1}}

\altaffiltext{1}{Dept. of Physics \& Astronomy, University 
of Wyoming, Laramie, WY 82071, USA}
\altaffiltext{2}{Steward Observatory, University 
of Arizona, Tucson, AZ 85721, USA}
\email{rsmullen@email.arionza.edu}

\begin{abstract}
We present multi-epoch spectroscopy of ``\hb,'' eccentric binaries with dynamic tidal distortions and tidally induced pulsations originally discovered with the \emph{Kepler} satellite.  Optical spectra of six known \hb\ using the Wyoming Infrared Observatory 2.3 m  telescope allow measurement of stellar effective temperatures and radial velocities from which we determine orbital parameters including the periods,  eccentricities, approximate mass ratios, and component masses.  These spectroscopic solutions confirm that the stars are members of eccentric binary systems  with eccentricities $e>0.34$ and periods $P=7$--20~days, strengthening conclusions from prior works which utilized purely photometric methods.  Heartbeat stars in this sample have A- or F-type primary components.  Constraints on orbital inclinations indicate that four of the six systems have minimum mass ratios $q=0.3$--0.5, implying that most secondaries are probable M dwarfs or earlier. One system is an eclipsing, double-lined spectroscopic binary with roughly equal-mass mid-A components ($q=0.95$), while another shows double-lined behavior only near periastron, indicating that the F0V primary has a G1V secondary ($q=0.65$). This work constitutes the first measurements of the masses of secondaries in a statistical sample of \hb. The good agreement between our spectroscopic orbital elements and those derived using a photometric model support the idea that photometric data are sufficient to derive reliable orbital parameters for \hb.
\end{abstract}


\section{Introduction}\label{intro}

In addition to being a planet hunter, the \emph{Kepler} satellite has uncovered a host of unexpected stellar phenomena, including new and exotic classes of binary stars.  Most stars form in binary stellar systems \citep{DandM1991}, making it imperative to understand their statistical properties since they expose the physical processes inherent in star formation and evolution. \emph{Kepler} data have revealed a new class of eccentric binaries in which the primary star undergoes tidal distortions and pulsations in response to perturbations from the secondary star.  An early example, KOI-54, was reported by \cite{Welsh2011} who observed a light curve that brightened by 7\% every 41.8~days and exhibited additional modulations between maxima. They attributed the primary brightening to the tidal interaction of a binary system near periastron, while the higher frequency variations were interpreted as tidally induced  oscillations.  These objects had been predicted by \cite{Witte1999}, \cite{Witte2002}, and \cite{Willems2002}, but \emph{ Kepler } allowed for an in-depth observational analysis of the tidal distortions.  Later works, such as \cite{Fuller2012}, \cite{Burkart2012}, and \cite{Oleary2014}, have shown that the dominant brightening near periastron passage is likely due to resonantly locked $m=2$ modes that trace the period.

\citet[hereafter TH12]{Thompson2012} reported the discovery of 17 additional such binary systems from the \emph{Kepler} archive; the tally is now over 130 \citep{Hambleton2013}. The multi-modal nature of the light curves resembles an electrocardiogram, earning them the name ``\hb''.  TH12 determined an orbital period from the light curve and used the model of tidal distortions presented in \cite{Kumar1995} to fit the eccentricity, angle of periastron, and orbital  inclination to the shape of each \emph{Kepler} light curve. While the models of TH12 appear to provide good fits to the data, their limited number of radial velocity measurements precluded a definitive measurement of the orbital parameters and could not address the velocity amplitudes and masses of the stellar components. 
  
In this paper, we present a three-year campaign of optical spectroscopy for a small sample of heartbeat stars.  We chose a sample of six stars from TH12 as a pilot survey to directly measure orbital solutions and test the accuracy of the photometry-based model fits to the orbital parameters. We confirm the binary nature of these systems and find them to be short-period, eccentric binaries. Thus, we provide an independent confirmation of the hypothesis that the unusual light curves of these systems stem from tidal interactions of eccentric binary star systems.  Our six targets were chosen to be representative of the sample of 17 systems presented in TH12, spanning the range of temperatures and periods while being bright enough to obtain high-signal-to-noise observations with the Wyoming Infrared Observatory 2.3 m telescope.  Section~\ref{sor} discusses our spectroscopic observations and data reduction methods.  Section~\ref{data} contains an analysis of radial velocities from the extracted spectra and a discussion of our independent temperature measurements. Section~\ref{orb} presents best-fit orbital solutions for the six systems and a comparison to the solutions from TH12.  All velocities reported in this work are in the Heliocentric frame of reference. 

\section{Spectroscopic Observations and Reductions }\label{sor}

\begin{deluxetable}{lcccc}
\tabletypesize{\scriptsize}
\centering
\tablewidth{0pt}
\tablecaption{Targets \label{obs.tab}}
\tablehead{
\colhead{Name} &\colhead{R. A. (J2000)} &\colhead{Dec. (J2000)} &\colhead{$Kp$ (mag)}&\colhead{\# Obs.}  }
\startdata           
\K1  & 19 20 27.025 & +38 23 59.46 &  9.002 & 23 \\
\K2  & 19 28 19.089 & +38 50 13.60 & 10.575 & 18 \\
\K3  & 19 10 24.766 & +39 23 51.63 & 12.159 & 21 \\
\K4  & 20 04 27.423 & +44 49 32.87 & 11.614 & 24 \\
\K5  & 19 40 38.801 & +46 45 02.23 & 10.886 & 16 \\
\K6  & 18 56 06.426 & +49 24 55.50 & 10.988 & 19 
\enddata
\end{deluxetable} 

Spectra were obtained at the Wyoming Infrared Observatory (WIRO) 2.3 m telescope over the period 2012 March through 2014 September and include 49 nights of observations with an average of 20 observations per star.  Table~\ref{obs.tab} provides a list of the targets, the right ascension and declination of each object, the $Kp$ magnitude (referring to the broad optical \emph{Kepler} bandpass, which corresponds roughly to the Cousins \emph{Rc} magnitude), and the number of nights that  each target was observed. The stars have $Kp$ magnitudes between 9.0 and 12.2. 

All data were obtained using the WIRO-Longslit Spectrograph with an e2v $2048\times2048$ CCD.  A 2000 line per millimeter grating in first order provided a spectral resolution of about 1.2~\AA\ FWHM in the center of the 5400--6700~\AA\ spectral range.  A 1{\arcsec}.2 $\times$ 110\arcsec\ slit oriented at a position angle of 0\deg\ yielded a reciprocal dispersion of 0.63~\AA\ per pixel and resolution R$\approx$4500 at 5800~\AA.   Exposures ranged from 600 s to 1800 s in 200--300 s increments, depending on the target brightness and nightly conditions.  Seeing ranged from about 1{\arcsec}.2 to 3\arcsec\ FWHM.

Reductions followed standard techniques for longslit spectroscopy.  Spectra were flat-fielded from dome quartz lamp exposures and wavelength calibrated using copper-argon arc lamp exposures to a rms of 0.02~\AA\ (1.0~\kms\ at 5800~\AA).  Exposures for a given star on a given night, taken sequentially in the observing sequence, were combined  to yield one spectrum with signal-to-noise ratio (SNR) of over 70 per 0.63~\AA\ pixel. All spectra were Doppler corrected to the Heliocentric reference frame.  The mean uncertainty on the night-to-night velocity zero point was determined to be 4.1~\kms\ rms based on extensive observations of OB stars within the Cygnus OB2 Association obtained at WIRO over the same time period \citep{Kobulnicky2014}.\footnote{ This uncertainty is a conservative estimate of the velocity uncertainties introduced by temperature changes and flexure in the Cassegrain spectrograph.}  This uncertainty was added in quadrature to the radial velocity uncertainty determined during cross correlation, discussed in Section~\ref{ccm}. For the system with the largest velocity amplitude, \K1, we expect the radial velocity to change by roughly 6~\kms\ per hour near periastron.  As the time for a series of exposures for this system is 10--15 minutes, we would expect the total change in radial velocity to be less than 3~\kms. Because this is significantly less than our other velocity uncertainties, we consider it negligible.

\begin{figure}
\epsscale{1.}
\centering
\plotone{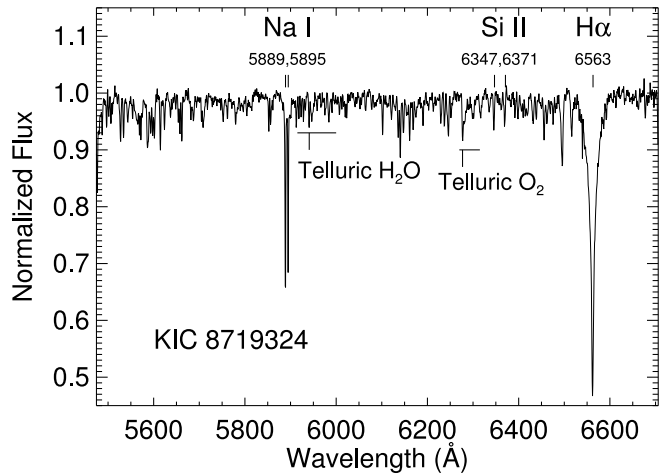}
\caption{Spectrum of \K4\ ($T_{\textrm{\tiny eff}}=7597$~K)as a representative example of a heartbeat star.  Labels indicate the strongest features in the spectrum, including H$\alpha$, stellar and interstellar \ion{Na}{1}~D $\lambda\lambda$5889,5895, \ion{Si}{2} $\lambda$6347~\AA, and \ion{Si}{2} $\lambda$6371~\AA. Horizontal bars mark telluric O$_2$ and H$_2$O bands.  Many other unlabeled metal lines are visible in the spectrum, providing significant cross-correlation power for determining radial velocities.
\label{spec}}
\end{figure}

Figure~\ref{spec} shows a continuum-normalized spectrum of \K4 (T$_{\textrm{\tiny eff}}=$7600~K) as a representative example of a heartbeat star. The strongest features in the spectrum are H$\alpha$ and \ion{Na}{1}~D, which contains both stellar and interstellar components.  Additional weak features from low-ionization metals are abundant in the spectrum and not individually labeled, with the exception of \ion{Si}{2} at $\lambda$6347 and $\lambda$6371~\AA.  No telluric corrections have been applied, so atmospheric features are present in the spectra, principally the O$_2$ band around 6300~\AA\ and the  H$_2$O band around 5900~\AA.

\section{Data Analysis}\label{data}

\subsection{Cross correlation Methods}\label{ccm}

We used the IRAF task XCSAO\footnote{XCSAO is a radial velocity analysis task from the Smithsonian Astrophysical Observatory package RVSAO.} to measure radial velocities from the spectra using cross-correlation techniques. The XCSAO task allows the user to specify a template spectrum against which to perform the cross-correlation.  Initially, we tried to use the highest signal-to-noise spectrum for each star in our sample as the template spectrum.  However, even our best spectra had a signal-to-noise ratio just over 100 so the observational velocity uncertainties were on the order of 1--10~\kms.  As some of these systems, such as \K5, have velocity amplitudes of $\sim$10~\kms, such large uncertainties preclude secure orbital solutions.  We then tried performing the cross-correlation against synthetic spectra from the MARCS \citep{Gustafsson2008} model stellar atmospheres of the nearest available temperature and gravity to those reported in TH12.\footnote{\K2 was correlated against a template of 7500~K with $\log g=3.5$; \K3 against a template of 6750~K with $\log g=4.5$; \K4 against a template of 7500~K with $\log g=3.5$; \K5 against a template of 7500~K with $\log g=3.5$; \K6 against a template of 6750~K with $\log g=4.5$.}  We corrected the vacuum wavelengths of the synthetic spectra to wavelengths in air and smoothed them by 1.1~\AA\ to match the resolution of our spectra. To ensure that resulting velocities were not sensitive to a specific template, we performed a test on target stars at 6750~K, 7500~K, and 8000~K in which we cross-correlated against templates mismatched in temperature by several hundred to a few thousand Kelvin.  The resulting velocities changed by $<0.2$~\kms\ and the uncertainties increased by ${<0.4}$~\kms. Similarly, we rotationally broadened a template by 100~\kms\ and found that the average change in the measured velocity was less than 1.5~\kms, much less than other velocity uncertainties inherent in our spectra. We conclude that the derived radial velocities are not sensitive to small mismatches in temperature, surface gravity, or rotational broadening between the star and template.  With these synthetic spectra, we can both estimate a systemic velocity and obtain average velocity {\it precisions} of $<2$~\kms.  The {\it accuracy} of the velocities is ultimately limited by the 4.1~\kms\ night-to-night uncertainties on the velocity zero point.  

When correlating the stellar velocities with XCSAO, the user can  specify specific regions of the spectrum to be ignored in the correlation.  It is particularly important, for example, to exclude the \ion{Na}{1}~D region because it contains both stellar and interstellar components. We considered four different combinations of spectral coverage in the cross-correlation. Our first cross-correlation method did not exclude any lines in the spectrum and was deemed to yield an unreliable velocity because the of the presence of the interstellar component of the \ion{Na}{1}~D lines.  The second method used only the H$\alpha$ (6530--6585~\AA) portion of the spectrum and was also discarded as sub-optimal because of the broad wings of H$\alpha$ leading to large velocity uncertainties. Our third method removed both the sodium (5883--5903~\AA) and H$\alpha$ features from the spectrum; the errors on the resulting velocities became larger in the absence of the strong H$\alpha$ line. Our adopted method for measuring velocities removed only the \ion{Na}{1}~D region from the spectrum and used the cross-correlation of the plethora of metal lines and the strong H$\alpha$.  The system \K2 shows some line splitting around periastron passage, so a measurement of the second peak in the cross correlation gave the velocity of the secondary when splitting was evident.  For the SB2 system \K1, whose components are nearly equal and therefore difficult to unentangle in the cross correlation spectrum, we instead measured the velocities of both components using a double Gaussian fit to the \ion{Si}{2} $\lambda$6347 line. The width and depth of the Gaussian fits were constrained using unblended spectra obtained near quadrature phase.  The velocities were not measured using the cross-correlation described above because of the difficulty separating the two peaks of the cross-correlation function when the components were partially blended.  Typical velocity uncertainties using this approach were 3--7~\kms, comparable to or larger than the night-to-night velocity zero point uncertainties.  

Tables~\ref{vel1.tab} and~\ref{vel2.tab} provide the ephemerides for the four SB1 systems \K3, \K4, \K5, and \K6, including the Heliocentric Julian date (HJD$-$2,400,000) of observation in column 1, the phase of the orbit in column 2, measured radial velocities ($V_{r1}$) with uncertainties for the primary in column 3, and the difference between observed and computed velocities from the orbital solution ($O_1-C_1$) for the primary in column 4. Table~\ref{velsb2.tab} provides the  provides the same information as Table~\ref{vel1.tab} for the SB2 systems \K2 and \K1, with the addition of $V_{r2}$ and $O_2-C_2$ for the secondary star in columns 5 and 6, respectively.

\begin{deluxetable}{lcrr}
\centering
\tabletypesize{\tiny}
\tabletypesize{\scriptsize}
\tablewidth{0pc}
\tablecaption{Ephemerides for Single-Lined Binary Systems \label{vel1.tab}}
\tablehead{
\colhead{} & 
\colhead{} & 
\colhead{$V_{r1}$} &
\colhead{$O_1-C_1$} \\
\colhead{HJD$-$2,400,000} &
\colhead{$\phi$} &
\colhead{(\kms)} &
\colhead{(\kms)}}
\tablecolumns{4}
\startdata  
\cutinhead{\K3}
56011.668 	 & 0.32 	 & 7.8(4.8) 	 & $-$3.7 \\
56015.633 	 & 0.78 	 & $-$27.9(4.5)  & $-$1.1 \\
56097.702 	 & 0.27 	 & 18.7(4.2) 	 & $-$0.9 \\
56098.721 	 & 0.39 	 & 4.7(4.2) 	 & 3.6 \\
56103.801 	 & 0.98 	 & 24.4(4.2) 	 & $-$2.9 \\
56107.761 	 & 0.44 	 & $-$4.0(4.2) 	 & 0.9 \\
56118.679 	 & 0.70 	 & $-$28.4(4.2)  & $-$1.6 \\
56128.816 	 & 0.87 	 & $-$8.3(4.3) 	 & 7.2 \\
56169.687 	 & 0.60 	 & $-$16.6(4.3)  & 4.5 \\
56178.756 	 & 0.65 	 & $-$18.9(4.2)  & 5.6 \\
56179.653 	 & 0.75 	 & $-$33.0(4.2)  & $-$5.5 \\
56435.801 	 & 0.38 	 & $-$2.3(4.2) 	 & $-$4.4 \\
56436.688 	 & 0.49 	 & $-$12.1(4.3)  & $-$1.6 \\
56437.682 	 & 0.60 	 & $-$21.3(4.3)  & $-$0.1 \\
56438.766 	 & 0.73 	 & $-$31.2(4.2)  & $-$3.7 \\
56445.718 	 & 0.53 	 & $-$14.0(4.3)  & 1.1 \\
56447.720 	 & 0.76 	 & $-$30.2(4.2)  & $-$2.8 \\
56796.706 	 & 0.13 	 & 46.7(4.3) 	 & $-$0.3 \\
56798.739 	 & 0.37 	 & 4.6(4.3) 	 & 0.1 \\
56799.812 	 & 0.49 	 & $-$8.4(4.2) 	 & 2.6 \\
56804.848 	 & 0.07 	 & 55.4(4.2) 	 & 2.4 \\
\cutinhead{\K4}
56011.727 	 & 0.53 	 & $-$28.5(12.5) & $-$2.2 \\
56015.707 	 & 0.92 	 & $-$15.5(4.5)  & 2.7 \\
56097.736 	 & 0.93 	 & $-$15.1(4.3)  & $-$2.0 \\
56098.747 	 & 0.03 	 & 37.1(4.3) 	 & $-$14.9 \\
56103.825 	 & 0.52 	 & $-$32.3(4.3)  & $-$6.0 \\
56107.765 	 & 0.91 	 & $-$19.7(4.3)  & 0.0 \\
56118.702 	 & 0.98 	 & 18.0(4.3) 	 & $-$7.7 \\
56128.760 	 & 0.96 	 & 18.0(5.9) 	 & 11.1 \\
56169.713 	 & 0.96 	 & 3.9(4.5) 	 & $-$4.2 \\
56178.780 	 & 0.85 	 & $-$16.2(4.3)  & 12.3 \\
56179.640 	 & 0.93 	 & $-$13.2(4.3)  & $-$1.1 \\
56438.751 	 & 0.25 	 & $-$10.4(4.3)  & $-$0.8 \\
56440.728 	 & 0.44 	 & $-$23.9(4.3)  & $-$1.1 \\
56441.769 	 & 0.54 	 & $-$31.2(4.3)  & $-$4.3 \\
56444.731 	 & 0.83 	 & $-$35.5(4.3)  & $-$6.1 \\
56445.731 	 & 0.93 	 & $-$9.5(4.3) 	 & 3.3 \\
56447.748 	 & 0.13 	 & 14.6(4.3) 	 & 5.0 \\
56464.690 	 & 0.78 	 & $-$34.8(4.5)  & $-$4.0 \\
56468.933 	 & 0.20 	 & $-$4.8(4.8) 	 & $-$1.3 \\
56795.870 	 & 0.14 	 & 10.5(4.5) 	 & 4.0 \\
56796.737 	 & 0.23 	 & $-$5.7(4.8) 	 & 1.4 \\
56798.805 	 & 0.43 	 & $-$15.0(5.0)  & 7.2 \\
56799.801 	 & 0.52 	 & $-$27.2(4.3)  & $-$1.0 \\
56804.792 	 & 0.01 	 & 69.5(4.3) 	 & 13.7 
\enddata 
\end{deluxetable}

\begin{deluxetable}{lcrr}
\centering
\tabletypesize{\tiny}
\tabletypesize{\scriptsize}
\tablewidth{0pc}
\tablecaption{Ephemerides for Single-Lined Binary Systems \label{vel2.tab}}
\tablehead{
\colhead{} & 
\colhead{} & 
\colhead{$V_{r1}$} &
\colhead{$O_1-C_1$} \\
\colhead{HJD$-$2,400,000} &
\colhead{$\phi$} &
\colhead{(\kms)} &
\colhead{(\kms)}}
\tablecolumns{4}
\startdata  
\cutinhead{\K5}
56439.692 	 & 0.93 	 & 5.2(4.8) 	 & $-$1.9 \\
56440.715 	 & 0.01 	 & 6.6(4.8) 	 & $-$0.2 \\
56442.727 	 & 0.18 	 & $-$1.0(4.5) 	 & 3.9 \\
56446.698 	 & 0.51 	 & $-$1.1(5.3) 	 & 2.2 \\
56461.671 	 & 0.76 	 & 4.3(4.8) 	 & 4.3 \\
56462.679 	 & 0.85 	 & 3.3(5.0) 	 & 0.9 \\
56795.801 	 & 0.64 	 & 0.4(9.1) 	 & 2.3 \\
56796.688 	 & 0.71 	 & $-$2.1(5.3) 	 & $-$1.1 \\
56798.772 	 & 0.88 	 & 1.0(4.8) 	 & $-$3.1 \\
56799.701 	 & 0.96 	 & 11.6(5.0) 	 & 1.1 \\
56804.774 	 & 0.39 	 & $-$9.4(5.0) 	 & $-$5.2 \\
56819.682 	 & 0.63 	 & $-$1.0(4.3) 	 & 1.1 \\
56821.706 	 & 0.80 	 & 0.4(4.8) 	 & $-$0.5 \\
56822.696 	 & 0.88 	 & 5.6(4.8) 	 & 1.8 \\
56825.688 	 & 0.13 	 & $-$5.4(4.3) 	 & $-$0.9 \\
56829.697 	 & 0.46 	 & $-$6.3(4.3) 	 & $-$2.7 \\
\cutinhead{\K6}
56097.661 	 & 0.88 	 & $-$39.4(4.2)  & $-$2.6 \\
56098.677 	 & 0.93 	 & $-$40.1(4.2)  & $-$0.2 \\
56107.735 	 & 0.41 	 & 1.9(4.2)	 &  0.3 \\
56118.653 	 & 0.99 	 & $-$29.7(4.2)  & $-$1.2 \\
56128.706 	 & 0.52 	 & $-$1.8(4.3)   &  3.1 \\
56169.727 	 & 0.68 	 & $-$18.6(4.3)  & $-$2.1 \\
56178.792 	 & 0.16 	 & $-$1.8(4.2)   & 15.8 \\
56179.625 	 & 0.20 	 & 11.9(4.2)	 & $-$0.8 \\
56436.675 	 & 0.75 	 & $-$22.7(4.2)  & $-$0.1 \\
56438.738 	 & 0.86 	 & $-$33.7(4.2)  &  0.5       \\
56439.678 	 & 0.91 	 & $-$39.4(4.2)  & $-$0.3 \\
56440.701 	 & 0.96 	 & $-$35.1(4.2)  &  2.1       \\
56444.686 	 & 0.17 	 & 12.5(4.2)	 & $-$1.2       \\
56445.688 	 & 0.22 	 & 11.7(4.2)	 & $-$0.1       \\
56447.763 	 & 0.33 	 & 3.7(4.2)	 & $-$2.4 \\
56796.859 	 & 0.73 	 & $-$20.0(4.2)  &  1.1       \\
56798.789 	 & 0.83 	 & $-$30.5(4.2)  &  1.1       \\
56799.721 	 & 0.88 	 & $-$36.6(4.2)  &  0.4       \\
56804.759 	 & 0.15 	 & 16.6(4.2)	 &  2.5       
\enddata 
\end{deluxetable}

\begin{deluxetable}{lcrrrr}
\centering
\tabletypesize{\tiny}
\tabletypesize{\footnotesize}
\tablewidth{0pc}
\tablecaption{Ephemerides for Double-Lined Binary Systems \label{velsb2.tab}}
\tablehead{
\colhead{} & 
\colhead{} & 
\colhead{$V_{r1}$} &
\colhead{$O_1-C_1$} &
\colhead{$V_{r2}$} &
\colhead{$O_2-C_2$} \\
\colhead{HJD$-$2,400,000} &
\colhead{$\phi$} &
\colhead{(\kms)} &
\colhead{(\kms)} &
\colhead{(\kms)} &
\colhead{(\kms)}}
\tablecolumns{6}
\startdata  
\cutinhead{\K2}
56439.706 	 & 0.32 	 & 15.4(4.3) 	 & $-$0.8 & \phd  & \phd \\
56440.687 	 & 0.37 	 & 17.2(4.3) 	 & $-$0.9 & \phd  & \phd \\
56442.709 	 & 0.47 	 & $-$5.3(4.3) 	 & $-$0.6 & \phd  & \phd \\
56446.711 	 & 0.66 	 & $-$42.3(4.3)  & $-$1.3 & \phd  & \phd \\
56458.683 	 & 0.25 	 & 8.4(4.3) 	 & $-$3.9 & \phd  & \phd \\
56460.665 	 & 0.35 	 & 12.9(4.3) 	 & $-$4.7 & \phd  & \phd \\
56795.770 	 & 0.84 	 & $-$16.3(4.5)  & 1.6    & \phd  & \phd \\
56796.846 	 & 0.89 	 & $-$9.4(4.2) 	 & 3.8    & \phd  & \phd \\
56798.826 	 & 0.99 	 & $-$5.1(4.3) 	 & 0.4    & \phd  & \phd \\
56799.734 	 & 0.03 	 & $-$1.9(4.3) 	 & 0.3    & \phd  & \phd \\
56804.870	 & 0.29 	 & 17.6(4.3) 	 & 3.3    & \phd  & \phd \\
56871.663 	 & 0.57 	 & $-$69.3(4.3)  & $-$5.1 & 67.8(35.8)  & $-$0.3 \\
56876.656 	 & 0.82 	 & $-$17.3(4.2)  & 2.6    & \phd  & \phd \\
56903.726 	 & 0.15 	 & 4.4(4.3) 	 & $-$1.2 & \phd  & \phd \\
56907.638 	 & 0.34 	 & 18.0(4.3) 	 & 0.7    & \phd  & \phd \\
56913.636 	 & 0.64 	 & $-$44.9(4.3)  & 1.4    & \phd  & \phd \\
56916.608 	 & 0.78 	 & $-$19.7(4.2)  & 3.9    & \phd  & \phd \\
56931.791 	 & 0.53 	 & $-$86.1(4.3)  & $-$3.1 & 95.3(8.7)  & 1.4  \\
\cutinhead{\K1}
56011.659 	 & 0.44 	 & $-$16.5(7.7)  & $-$5.7 	 & $-$19.0(25.0) & 1.8 \\
56015.601 	 & 1.00 	 & $-$90.5(5.9)  & $-$12.2 	 & 46.9(7.7) 	 & $-$4.2 \\
56097.685 	 & 0.64 	 & 19.7(4.5) 	 & $-$4.1 	 & $-$47.8(5.0)  & 9.9 \\
56098.703 	 & 0.79 	 & 44.3(4.8) 	 & $-$3.4 	 & $-$83.6(5.6)  & $-$0.5 \\
56103.792 	 & 0.51 	 & $-$7.3(7.1) 	 & $-$8.4 	 & $-$17.4(10.0) & 16.2 \\
56107.717 	 & 0.06 	 & $-$146.5(5.0) & $-$13.7 	 & 122.6(6.3) 	 & 13.4 \\
56118.671 	 & 0.62 	 & 6.9(5.3) 	 & $-$13.0 	 & $-$48.1(6.3)  & 5.4 \\
56128.722 	 & 0.04 	 & $-$116.6(6.7) & 21.3 	 & 94.8(8.3) 	 & $-$19.7 \\
56169.673 	 & 0.85 	 & 58.8(5.3) 	 & 2.0  	 & $-$92.6(6.3)  & 0.3 \\
56178.744 	 & 0.14 	 & $-$97.5(5.3)  & $-$2.0 	 & 95.7(6.7) 	 & 26.2 \\
56179.619 	 & 0.26 	 & $-$42.0(5.3)  & 9.1  	 & 36.8(6.3) 	 & 14.7 \\
56435.788 	 & 0.59 	 & 25.9(5.0) 	 & 10.6 	 & $-$40.5(6.3)  & 8.2 \\
56436.711 	 & 0.72 	 & 30.6(5.6) 	 & $-$6.5 	 & $-$51.8(7.1)  & 20.1 \\
56437.705 	 & 0.86 	 & 61.7(4.5) 	 & 3.7  	 & $-$84.7(5.0)  & 9.5 \\
56438.716 	 & 0.01 	 & $-$99.0(6.3)  & $-$6.0 	 & 61.0(8.3) 	 & $-$5.7 \\
56439.722 	 & 0.15 	 & $-$98.7(5.0)  & $-$8.3 	 & 62.3(5.9) 	 & $-$1.7 \\
56443.716 	 & 0.71 	 & 9.1(5.3) 	 & $-$26.9 	 & $-$100.0(6.3) & $-$29.3 \\
56444.714 	 & 0.86 	 & 52.3(5.9) 	 & $-$5.1 	 & $-$95.8(7.1)  & $-$2.2 \\
56445.709 	 & 1.00 	 & $-$70.1(4.8)  & $-$0.4 	 & 43.6(5.6) 	 & 1.7 \\
56446.952 	 & 0.17 	 & $-$81.4(5.3)  & $-$1.6 	 & 46.6(6.3) 	 & $-$6.1 \\
56447.698 	 & 0.28 	 & $-$46.0(5.3)  & 0.0  	 & 7.8(6.3) 	 & $-$9.0 \\
56467.678 	 & 0.11 	 & $-$110.1(5.0) & $-$2.8 	 & 83.0(5.9) 	 & 0.9 \\
56470.708 	 & 0.54 	 & 34.4(5.6) 	 & 27.1 	 & $-$32.6(4.8)  & 7.4 
\enddata 
\end{deluxetable}

\subsection{Temperatures}\label{temp}

\begin{figure}
\centering
\epsscale{01.}
\plotone{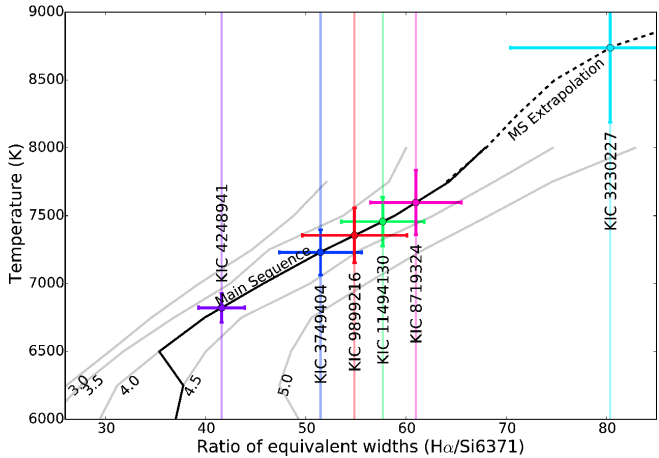}
\caption{Effective temperature versus ratio of equivalent widths EW$_{\mathrm{H}\alpha}$/EW$_{\mathrm{Si\textsc{ii}}}$.  Gray tracks show the measured equivalent width ratios for MARCS model atmospheres as a function of temperature for five different surface gravities, as labeled from $\log g=$3.0 to $\log g=$5.0. The solid black line shows the nominal main sequence, while the dashed line shows a quartic extrapolation of the 7500--7750~K points used to estimate higher temperatures.  Vertical lines mark the measured EW ratio of each star in our sample and provide a measure of temperature at various surface gravities.  The temperature of each heartbeat star is estimated using a linear interpolation along the main-sequence track, denoted by the circle and error bars.   (A color version of this figure is available in the online journal.)
\label{temps}} 
\end{figure}

To measure temperatures for the six systems, we used the MARCS model atmospheres \citep{Gustafsson2008} and smoothed the continuum-normalized model spectra to the resolution of the observed spectra.  We measured the equivalent widths of the \ion{Si}{2} $\lambda$6371.4 and H$\alpha$ lines for both observed and model spectra.  The EW of the silicon line was measured by fitting it with a Gaussian profile, while the EW of H$\alpha$ was measured by fitting it with a superposition of two Lorentzian profiles.  We measured the equivalent widths (EW) for the range of MARCS model atmospheres between temperatures of 6000~K and 8000~K and $\log g$ between 3.0 and 5.0.  Figure~\ref{temps} plots the stellar effective temperature versus the ratio of equivalent widths EW$_{\mathrm{H}\alpha}$ / EW$_{\mathrm{Si\textsc{ii}}}$  for both the MARCS models and our data. The gray tracks illustrate the relationship between temperature and EW ratio for five different model surface gravities, as labeled, with the black track showing the nominal main sequence in this temperature range.\footnote{We adopt the relation between spectral type and $\log g$ recommended by \cite{Castelli04}. For late-A through late-F stars typical of our sample, $\log g$ is $\sim$4.0. }  We also plot a quartic extrapolation of the last several points on the main sequence track to provide an estimate of EW ratios for hotter stars. Each  vertical line marks the equivalent width ratio of a star in our sample.  Symbols with error bars on the corresponding vertical line lie at the intersection of the vertical line with the main sequence. The uncertainties on the equivalent width ratio are measured from the error in the profile fit to the spectral lines; uncertainties on temperature are derived from the uncertainties of the EW ratio and the local slope of the adopted main-sequence track.  Although our data are not of sufficient resolution and wavelength coverage to calculate a $\log g$ for each star, other measurements, such as the photometric measurements in the \emph{Kepler} Mikulski Archive for Space Telescopes (MAST) archive, suggest that these stars have $\log g=4.0\pm0.2$.  As our grid of $\log g$ values is much coarser than this, we assume a main sequence luminosity class for these objects.  While this assumption may have implications on the temperature, it will have minimal impact on the mass.  Thus, we advocate that our temperatures be used with caution, but find them sufficient for the present analysis.  Most stars fall securely in the range $T_{\textrm{\tiny eff}}=6700$--8100~K and are well-constrained, within the typical uncertainties of $<250$~K, by the EW ratio measurement.  One star (\K1) has EW ratios that indicate higher temperatures than the ranges covered by the MARCS models, so we have estimated the temperature by extrapolating (dashed line) the 7500--7750~K main-sequence points.  Tables~\ref{solutions1.tab}  (line 10) and \ref{solutions2.tab} (line 11) list our best-fit temperatures and uncertainties along with the temperatures estimated by TH12. Our temperature measurements typically lie within 1--2$\sigma$ of those reported by TH12.  \K6 is the most discrepant; TH12 report a temperature of 6750~K, while we find a temperature of closer to 7500~K.  Our EW ratio temperature estimate for \K1 is particularly uncertain because of the double-lined nature of the system (blended lines) and the fact that \K1's temperature lies in the extrapolated regime. Consequently, we prefer a different method to estimate temperature as discussed in section~\ref{k1t}. 

\section{Orbital Solutions}\label{orb}

We used the power spectrum generated from a discrete Fourier transform of the radial velocity data to make an initial estimate of the orbital period for each system.  \cite{Kiminki2012} provide an explanation of this method applied to massive O and B type stars. The power spectra for \hb\ are more complex than those for more ``normal'' binary systems that do not exhibit tidally induced pulsations.  Power spectra for  \hb\ show several strong peaks at periodicities of a few days to a few tens of days.  The strongest peak is typically $>$50\% higher than other peaks and, in all cases, correctly selects the orbital period of the system.  These periods agree well with the more precise periods inferred by TH12 from the the \emph{Kepler} light curves, lending confidence to this independent determination of the orbital period.   Other strong peaks (on the order of 2--4~days) likely correspond to the harmonic modes of the primary star induced by the close passage of the secondary star, as argued by TH12.  We used the Binary Star Combined Solution Package \citep{Gudehus2001} to solve for the orbital parameters and uncertainties, including a more refined period.  Table~\ref{solutions1.tab} displays our best-fit results and the fitted parameters from TH12 for each SB1, listing the period ($P$) in days, the eccentricity of the system ($e$), the angle from node to periastron in degrees ($\omega$), the systemic velocity in~\kms\ ($\gamma$), the epoch of periastron ($T_0$), the projected velocity semi-amplitude for the primary in~\kms ($K_1$),  the projected semi-major axis of the system in AU ($a\sin i$), the mass function ${(M_2\sin i)^3}/{(M_1+M_2)^2}$ in~\msun, the estimated inclination range in degrees ($i$), the estimated effective temperature of the primary ($T_{\textrm{\tiny eff}}$), the estimated spectral type and adopted mass of the primary ($M_1$), the inferred secondary mass at the maximum estimated inclination ($M_2$), the minimum mass ratio ($q$; calculated as $M_2$ at $i_{\textrm{\tiny max}}$ divided by $M_1$), and the rms of the residuals of the best-fitting solution to the orbital elements in \kms\ ($\textrm{rms}_{1}$). The first eight rows of the table are directly fitted with the \cite{Gudehus2001} package while the next seven are estimated using methods described above and below.  Table~\ref{solutions2.tab} lists the best-fit orbital elements and the TH12 fit for the SB2 systems, with the period ($P$) in days, the eccentricity of the system ($e$), the angle from node to periastron in degrees ($\omega$), the systemic velocity in~\kms\ ($\gamma$), the epoch of periastron ($T_0$), the projected velocity semi-amplitude for the primary ($K_1$) and secondary ($K_2$) in~\kms, the projected semi-major axis of the system in AU ($a\sin i$), the mass ratio $q$, the sum of the masses $(M_1+M_2)\sin^3 i$ (\msun), the estimated inclination range in degrees ($i$), the estimated effective temperature of the primary ($T_{\textrm{\tiny eff}}$), the estimated spectral type and adopted mass of the primary ($M_1$) and secondary ($M_2$), and the rms velocity of the residuals of the best-fitting solution to the primary and secondary ($\textrm{rms}_{1}$ and $\textrm{rms}_{2}$, respectively).

\begin{deluxetable*}{lrlrlrlrl}
\centering
\tabletypesize{\tiny}
\tablewidth{0pt}
\tablecaption{Orbital Elements of Single-Lined Binaries \label{solutions1.tab}}
\tablehead{
\colhead{Element}  &
\multicolumn{2}{c}{\K3 } &
\multicolumn{2}{c}{\K4 } &
\multicolumn{2}{c}{\K5 } &
\multicolumn{2}{c}{\K6 } \\
\colhead{} &
\colhead{This work} &
\colhead{TH12}      &
\colhead{This work} &
\colhead{TH12}      &
\colhead{This work} &
\colhead{TH12}      &
\colhead{This work} &
\colhead{TH12}}
\startdata            
$P$ (days)               & $8.645\pm0.002$  & $8.645\pm0.001$	 & $10.235\pm0.005$ & $10.233\pm0.001$   & $11.987\pm0.018$ & $10.916\pm0.001$   & $18.973\pm0.008$& $18.956\pm0.001$	 \\
$e$                      & $0.34\pm0.04$    & $0.42\pm0.01$	 & $0.64\pm0.05$    & $0.60\pm0.01$	 & $0.66\pm0.24 $   & $0.65\pm0.01$	 & $0.49\pm0.05$   & $0.62\pm0.01$	 \\
$\omega$ (deg)           & $308\pm5$	    & $310\pm4$ 	 & $330\pm7$	    & $343\pm1$ 	 & $35\pm16$	    & $81\pm1$  	 & $70\pm4$	   & $62\pm1$		 \\
$\gamma$ (\kms)          & $4.4\pm1.1$      & \phd		 & $-11.7\pm1.9$    & \phd		 & $-1.0\pm1.0$     & \phd		 & $-8.6\pm0.7$    & \phd		 \\
$T_0$ (HJD$-$2,450,000)  & $6095.3\pm0.1$   & \phd		 & $6292.9\pm0.1$   & \phd		 & $6296.7\pm0.4$   & \phd		 & $6289.7\pm0.1$  & \phd		 \\
$K_{1}$ (\kms)           & $40.3\pm1.8$     & \phd		 & $43.5\pm3.1$     & \phd		 & $8.5\pm4.6$      & \phd		 & $27.0\pm1.0$    & \phd		 \\
$a \sin i$ (AU)          & $\sim$0.03	    & \phd		 & $\sim$0.03	    & \phd		 & $\sim$0.01	    & \phd		 & $\sim$0.04	   & \phd		 \\
Mass function (\msun)	 &$0.051\pm0.006$   &\phd 		 & $0.065\pm0.013$  & \phd	  	 & $0.0003\pm0.0002$  & \phd		 & $0.026\pm0.002$ & \phd		 \\
$i$  (degrees)           & 31--64	    & $68\pm6$	  	 & $\ge37$	    & $76\pm1$  	 & 6--15	    & $21\pm1$  	 & 23--67	   & $48\pm1$		 \\
$T_{\textrm{\tiny eff}}$ (K) & $6822\pm107$ & 6750		 & $7597\pm236$     & 7750		 & $7355\pm201$     & 7500		 & $7456\pm179$    & 6750		 \\
Estimated S.~C.$_1$      & F3V  	    & \phd		 & A8V  	    & \phd		 & F0V  	    & \phd		 & A9V  	   & \phd		 \\
$M_{1}$ (adopted; \msun) & 1.5  	    & \phd		 & 1.7  	    & \phd		 & 1.6  	    & \phd		 & 1.7  	   & \phd		 \\ 
$M_{2}$ at $i_{\textrm{\tiny max}}$ (\msun) & 0.7 & \phd	 & 0.7  	    & \phd		 & 0.5  	    & \phd		 & 0.6  	   & \phd		 \\ 
Minimum $q$              & 0.46  	    & \phd		 & 0.43 	    & \phd		 & 0.28 	    & \phd		 & 0.33 	   & \phd		 \\
$\textrm{rms}_1$ (\kms)  & 3.33 	    & \phd		 & 6.44 	    & \phd		 & 2.50 	    & \phd		 & 3.84 	   & \phd		 
\enddata
\end{deluxetable*} 

\begin{deluxetable*}{lrlrl}
\centering
\tabletypesize{\tiny}
\tablewidth{0pt}
\tablecaption{Orbital Elements of Double-Lined Binaries \label{solutions2.tab}}
\tablehead{
\colhead{Element}  &
\multicolumn{2}{c}{\K1 } &
\multicolumn{2}{c}{\K2 } \\
\colhead{} &
\colhead{This work} &
\colhead{TH12}      &
\colhead{This work} &
\colhead{TH12}}
\startdata            
$P$ (days)               & $7.051\pm0.001$  & $7.047\pm0.001$   & $20.316\pm0.010$ & $20.307\pm0.001$  	\\
$e$                      & $0.60\pm0.04$    & $0.59\pm0.01$     & $0.66\pm0.03$	   & $0.64\pm0.01$     	\\
$\omega$ (deg)           & $293\pm4$	    & $292\pm1.2$       & $308\pm3$	   & $302\pm1$      	\\
$\gamma$ (\kms)          & $-15.7\pm1.7$    & \phd    		& $-13.9\pm0.8$    & \phd    		\\
$T_0$ (HJD$-$2,450,000)  & $6311.76\pm0.03$ & \phd    		& $6656.85\pm0.10$ & \phd     		\\
$K_{1}$ (\kms)           & $98.5\pm5.4$     & \phd   	 	& $50.0\pm2.1$	   & \phd     		\\
$K_{2}$ (\kms)           & $104.9\pm6.1$    & \phd    		& $78.1\pm5.7$	   & \phd     		\\
$a \sin i$ (AU)          & $\sim$0.05	    & \phd    		& $\sim$0.07	   & \phd     		\\
$q$                      & $0.95\pm0.05$    & \phd    		& $0.64\pm0.09$    & \phd     		\\
$(M_1+M_2)\sin^3 i$ (\msun)& $3.1\pm0.3$    & \phd		& $1.9\pm0.3$	   & \phd		\\
$i$  (degrees)           & 66--71	    & $43\pm1$      	& 51--65 	   & $38\pm1$  		\\
$T_{\textrm{\tiny eff}}$ (K) & $\sim$8000   & 8750              & $7231\pm166$     & 7500     		\\
Estimated S.~C.$_1$      & A6V              & \phd    		& F0V		   & \phd     		\\
$M_{1}$ (adopted; \msun) & 2.0              & \phd    		& 1.6		   & \phd     		\\ 
Estimated S.~C.$_2$      & A6V	            & \phd   		& G1V		   & \phd     		\\
$M_{2}$ (\msun)          & 2.0              & \phd    		& 1.0 		   & \phd     		\\ 
$\textrm{rms}_{1}$ (\kms) & 12.11    	    & \phd   		& 2.01		   & \phd     		\\
$\textrm{rms}_{2}$ (\kms) & 11.14    	    & \phd   		& 0.85		   & \phd     		
\enddata
\end{deluxetable*} 

\subsection{Inclinations}

While SB1 radial velocities alone cannot constrain system inclination, we can put limits on the inclinations of these systems and consequently place limits on the stellar masses. These masses and inclinations depend  upon the adopted primary mass, an assumed stellar mass-radius relationship, and the accuracy of our measured orbital elements.  There are two key pieces of information that allow inclination constraints: first, the mass of the secondary cannot be larger than the mass of the primary, and second, these systems have been observed with \emph{Kepler} to have either eclipsing or non-eclipsing behavior.  The first criterion provides an upper limit on secondary mass and lower limit on the inclination.  If a system is eclipsing, we can put a more stringent lower limit on the inclination by calculating the impact parameter (the projected distance at conjunction relative to the observer), given by \[b=\frac{a\cos i}{R_{\textrm{\tiny Primary}}} \left(\frac{1-e^2}{1+e\sin\omega}\right)\] \citep{Winn2010}. Here, $b=1$ indicates where one would begin to observe a grazing eclipse. If a system is not eclipsing, we  estimate an upper limit for the inclination and a lower limit on the mass based on the inclination at which we would expect the system to show an eclipse. We calculate the impact parameter by adopting a main sequence dwarf mass consistent with our observed temperature and then applying a stellar mass-radius relationship.  We examined mass-radius relations of both  \cite{Demircan1991} and \cite{Malkov2007}, and there is excellent agreement in the calculated masses and inclinations between the two prescriptions (less than 1.5\deg\ in inclination and 0.04~\msun\ in mass).  Tables~\ref{solutions1.tab}~and~\ref{solutions2.tab} and the sections below contain the individual results for each star.

\subsection{\K3}\label{k3}

\begin{figure}
\centering
\epsscale{1.}
\plotone{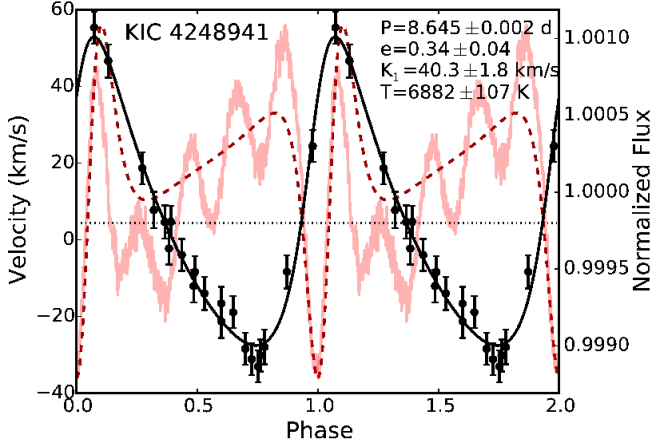}
\caption{The radial velocity data and best-fit solution for the 8.645 day period SB1 system \K3.   Circles denote the primary star's velocity.  The left vertical axis shows Heliocentric radial velocity, and the horizontal axis shows the orbital phase.   The highly structured curve plotted in the background is the \emph{Kepler} light curve with the normalized flux on the right vertical axis. A model light curve created with the orbital elements from the data in this paper and the TH12 inclination is shown in the dashed curve. (A color version of this figure is available in the online journal.)
\label{K424soln}} 
\end{figure}

We observed the SB1 \K3 on 21 nights from 2012 March through 2014 May. Our temperature estimate of $T=6822\pm107$~K agrees with TH12's temperature of 6750~K. We find a best-fit period of $P=8.645\pm0.002$~days, which corresponds exactly to the TH12 period.  The eccentricity is fit as $e=0.34\pm0.04$, which is similar to TH12's reported eccentricity of 0.42.  We also find a systemic velocity of $\gamma=4.4\pm1.1$~\kms\ and a velocity amplitude of $K_1=40.3\pm1.8$~\kms.  This system has a projected semi-major axis of $a\sin i=0.03$~AU. Figure~\ref{K424soln} displays the best-fit orbital solution, phased velocity curve, and phased light curve for this system. Black circles denote the radial velocity measurements obtained with WIRO, and the solid black line shows our best-fit solution.  The highly structured curve plotted in the background shows the phased light curve taken from the publicly released simple aperture photometry (SAP) data in the \emph{Kepler} MAST Archive, which was then phased and normalized with PyKE \citep{Still2012}. The overplotted dashed curve shows the \cite{Kumar1995} light curve model generated using our orbital parameters and the TH12 inclination.\footnote{The model of photometric tidal variations was taken from equation 44 in \cite{Kumar1995}, presented here as written in  equation 1 of TH12: \[\frac{\delta f}{f}=S\cdot\frac{1-3\sin^2{i} \sin^2\left( \phi(t)-\omega \right)}{\left( R(t)/a \right)^3 } + C\] Here, $\frac{\delta f}{f}$ is the change in flux of the light curve, $S$ is a scaling factor to match the amplitude of the normalized light curve, $i$ is the system inclination (for our purposes, taken from TH12), $\phi(t)$ is the true anomaly (calculated by first computing the eccentric anomaly, which itself is dependent upon the eccentricity), $\omega$ is the angle from node to periastron, $R(t)$ is the time-variable separation between the two stars (dependent upon $a$, $e$, and the eccentric anomaly), $a$ is the semi-major axis, and $C$ is an offset to represent the zero-point of the normalized light curve.} The horizontal axis shows the orbital phase; the left vertical axis marks the velocity of the measured components in~\kms, while the right vertical axis shows the normalized flux of the light curve. Note that the point of maximum velocity amplitude (at phase=1) coincides with largest variation in the light curve.  This is consistent with the scenario proposed by TH12 in which the primary star undergoes large tidal distortions near the time of periastron. 

To estimate the mass of the secondary, we assume a primary spectral type of F3V (1.5~\msun) and no eclipses. We find that the inclination must lie in the range  31--64\deg, corresponding to a secondary masse of 1.5--0.7~\msun. Thus, the minimum mass ratio is $q=0.46$. TH12's inclination of 68$\pm$5.5\deg\ is consistent with our range.

\subsection{\K4}\label{k4}

\begin{figure}
\centering
\epsscale{1.0}
\plotone{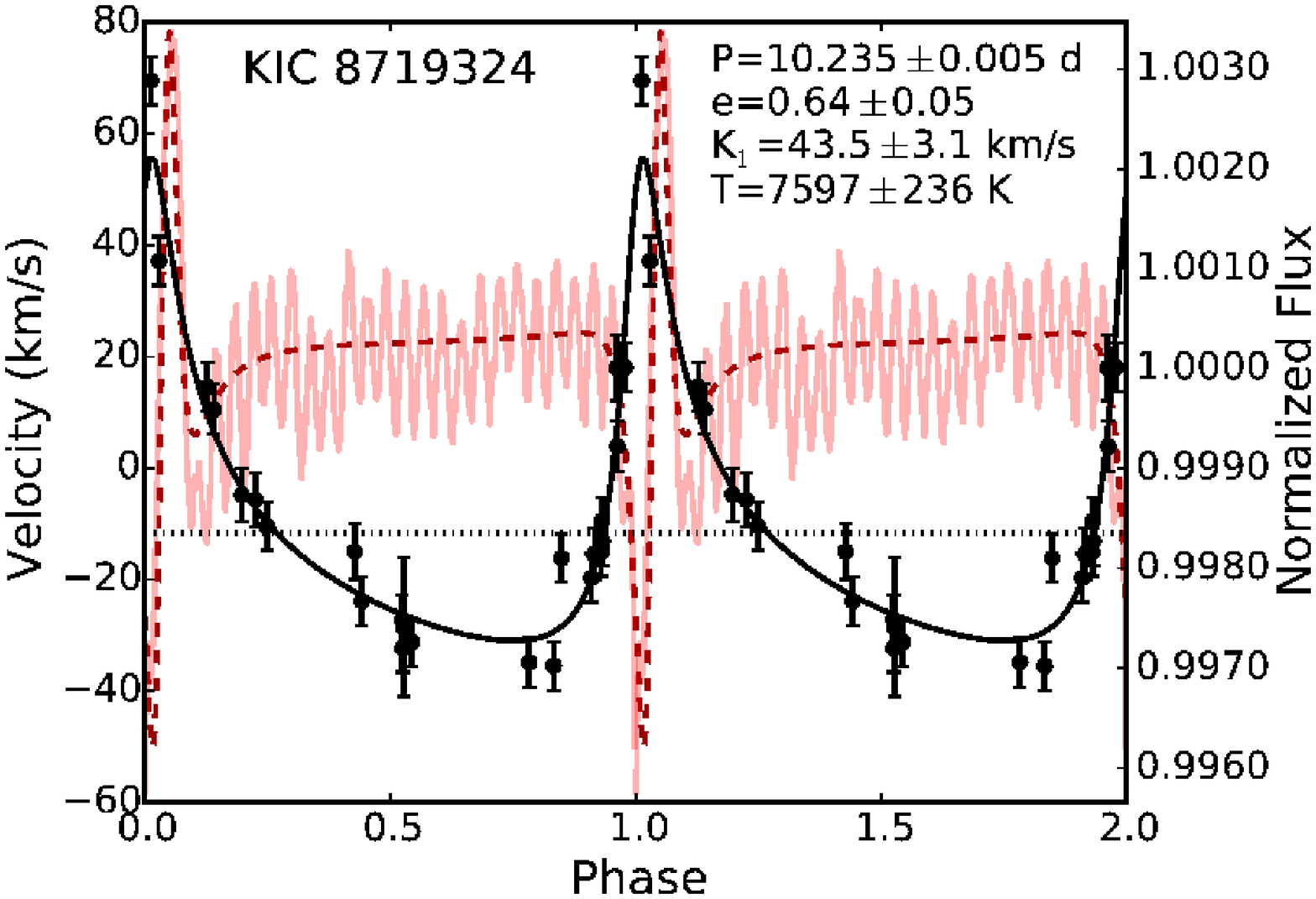}
\caption{The radial velocity data and best-fit solution for the 10.235 day period SB1 system \K4.  This plot uses the same conventions as Figure~\ref{K424soln}.  (A color version of this figure is available in the online journal.)
\label{K871soln}} 
\end{figure}

\K4, an SB1 system, was observed on 24 nights from 2012 March through 2014 May.   We find a temperature of $T=7597\pm236$~K, which is similar to TH12's value of 7750~K. This system shows a grazing eclipse signature in the light curve as reported in TH12, although the light curve is dominated by the  signature of the \K4 tidal modes.  We find a period of $P=10.235\pm0.005$~days, an eccentricity of $e=0.64\pm0.05$, a systemic velocity of $\gamma=-11.7\pm1.9$~\kms, a velocity amplitude of $K_1=43.5\pm3.1$~\kms, and a projected semi-major axis of $a\sin i=0.03$~AU.  Our solution agrees well with the TH12 period of 10.233~days and eccentricity of 0.60. Figure~\ref{K871soln} depicts the best-fit solution, data, and the \emph{Kepler} light curve for \K4.

We estimate the mass of the secondary to be a minimum of 0.7~\msun\ by assuming an A8V spectral type with mass 1.7~\msun\ for the primary star.  By using the presence of a grazing eclipse, we constrain the inclination to be greater than 37\deg, which limits the secondary mass to less than 1.5~\msun.  This inclination range supports the TH12 value of 76\deg\ and provides a minimum mass ratio of $q=0.43$ at $i=90$\deg.
 
\subsection{\K5}\label{k5}

\begin{figure}
\centering
\epsscale{1.0}
\plotone{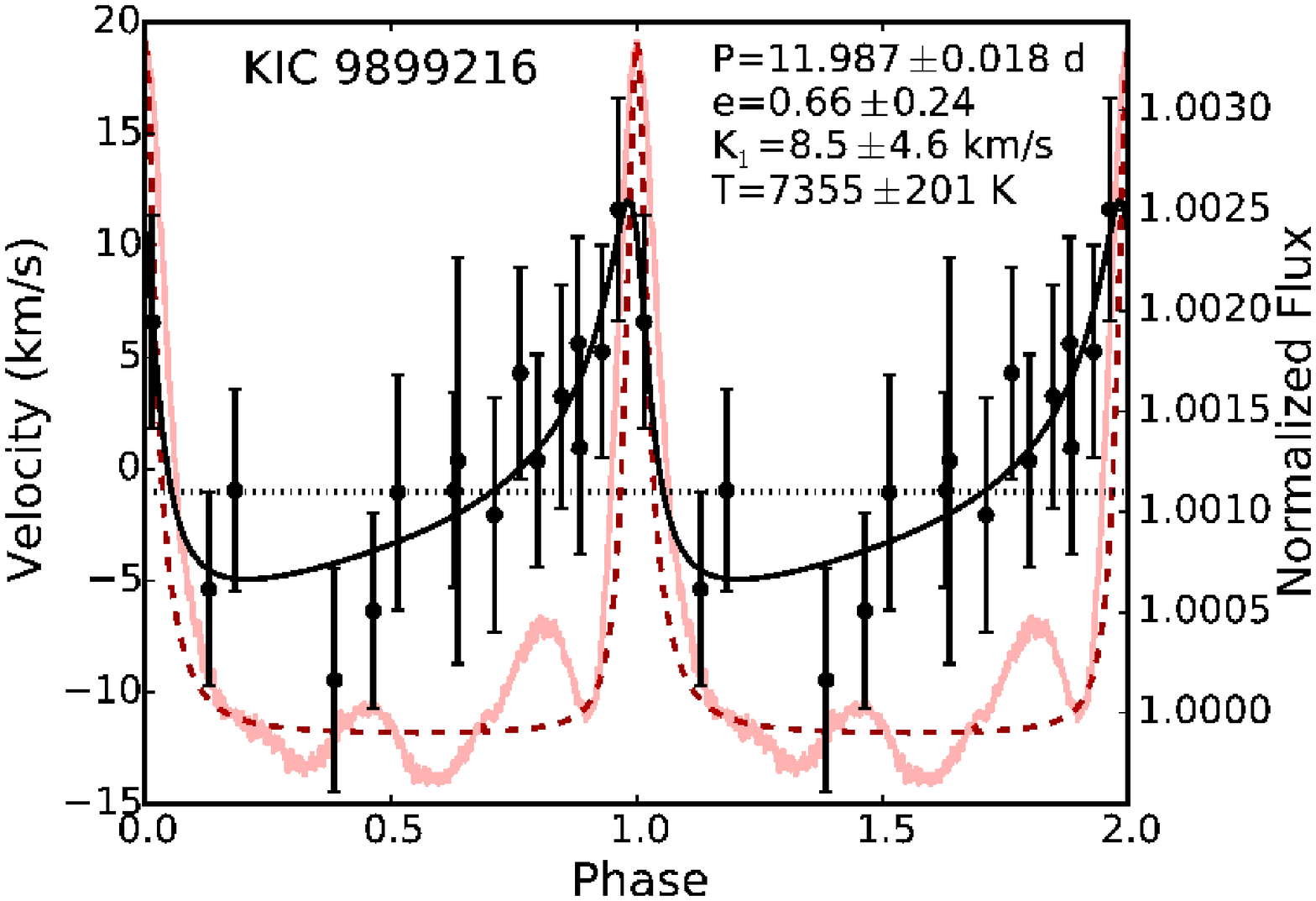}
\caption{The radial velocity data and best-fit solution for the 11.987 day period SB1 system \K5.  This plot uses the same conventions as  Figure~\ref{K424soln}.  (A color version of this figure is available in the online journal.)
\label{K989soln}} 
\end{figure}

\K5, another SB1, has an effective temperature of $T=7355\pm201$~K, in agreement with TH12's temperature of 7500~K. We have sixteen nights of observation for this object from 2013 May to 2014 June.  The spectroscopic period of this system is  $P=11.987\pm0.018$~days, which is somewhat larger than TH12's period of 10.916~days. This discrepancy in periods likely arises because of the low velocity amplitude of $K_1=8.5\pm4.6$~\kms, which is roughly twice the typical velocity uncertainty on each measurement. We did attempt to fix our orbital solution at the period of TH12, but this resulted in a much poorer solution.  Additional data  will be required to reconcile the difference in periods.  However, we do find the eccentricity of $e=0.66\pm0.24$ to be comparable to TH12's eccentricity of 0.65.  We also measure the systemic velocity to be $\gamma=-1.0\pm1.0$~\kms\ and the projected semi-major axis to be $a\sin i=0.01$~AU.  Figure~\ref{K989soln} depicts the best-fit solution, data, and \emph{Kepler} light curve for \K5.

TH12 estimate that \K5 has an inclination of 21\deg.  This is the lowest estimated inclination of our sample of \hb.  Our inclination range of 6--15\deg\ corroborates the TH12 result. The minimum secondary mass of 0.5~\msun, coupled with the estimation that \K5 is an F0V star with mass 1.6~\msun, gives a minimum mass ratio of $q=0.28$.

\subsection{\K6}\label{k6}

\begin{figure}
\centering
\epsscale{1.0}
\plotone{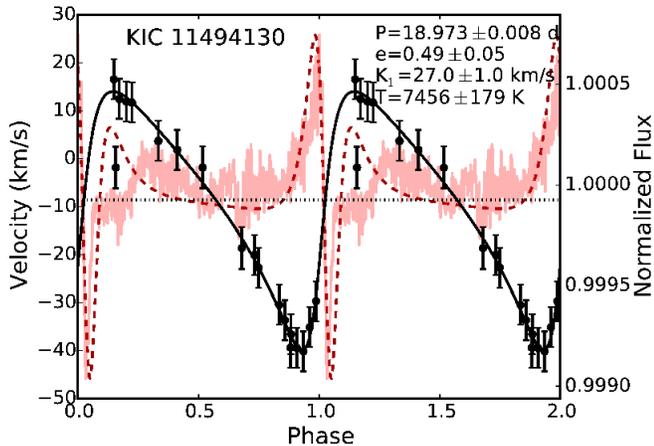}
\caption{The radial velocity data and best-fit solution for the 18.973 day period SB1 system \K6.  This plot uses the same conventions as  Figure~\ref{K424soln}. The discrepant data point at phase 0.16 is not included in the fit. 
\label{K1149soln}} 
\end{figure}

We gathered 19 nights of data on \K6 from 2012 June through 2014 May.  We find a temperature of $T=7456\pm179$~K, which contrasts greatly with TH12's temperature of 6750~K.  TH12 fit this star with a $\log g$ of 4.5; if we adopt the same value, our temperature would drop to near 7200~K.  A $\log g$ of 5.0 yields a temperature much closer to the TH12 value, but is significantly larger than the canonical values for 
a main sequence star of this temperature.  The best-fit orbital solution yields a period of $P=18.973\pm$0.008~days, which is similar to the TH12 period of 18.956~days.  The best-fit eccentricity is $e=0.49\pm0.05$, somewhat less than the TH12 eccentricity of 0.62. However, our best-fit eccentricity is plausible if TH12 underestimated the width of the leading wing of the light curve pulsation, which is possible with the significant variability near periastron.  We find a systemic velocity of $\gamma=-8.6\pm0.7$~\kms, a primary velocity amplitude of $K_1=27.0\pm1.0$~\kms, and a projected semi-major axis of $a\sin i=0.04$~AU.  Figure~\ref{K1149soln} displays the velocity data and best-fit solution, in addition to the \emph{Kepler} light curve.  The discrepant data point at phase 0.16 is not included in the fit. 

Depending on the actual temperature of the star, the mass of the primary will differ.  If we assume an A9V spectral type (7500~K and mass 1.7~\msun, from this work) for the primary, we find an inclination range of 23--67\deg.  This leads to a minimum secondary mass of 0.6~\msun\ and a minimum mass ratio of $q=0.33$.  Assuming instead an F5V spectral type for the primary (6700~K and mass 1.4~\msun, from TH12), we find an inclination range of 25--70\deg, giving a minimum secondary of mass 0.5~\msun.  However, this leads to the same minimum mass ratio.

\subsection{\K2}\label{k2}

\begin{figure}
\centering
\epsscale{1.0}
\plotone{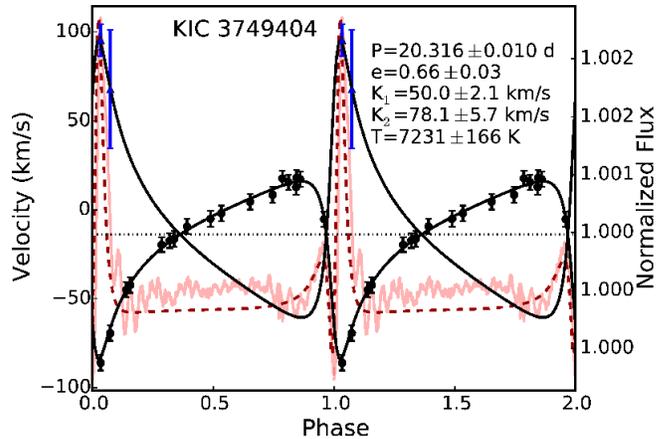}
\caption{Radial velocity data and best-fit solution for the 20.316 day period SB2 system \K2. Circles denote the primary, while triangles show data for the secondary component. This figure follows the same conventions as Figure~\ref{K424soln}.  (A color version of this figure is available in the online journal.)
\label{K374soln}} 
\end{figure}

\K2 is an SB2 for which we have eighteen observations between 2013 May to 2014 September.   We find an effective temperature of $T=7231\pm166$~K, which agrees well with TH12's reported temperature of 7500~K.  The best-fit period of $P=20.316\pm0.010$~days matches TH12's photometric period of 20.307~days.  The eccentricity was best fit with a value of $e=0.66\pm0.03$, consistent with the TH12 estimate of 0.64. Our fit also produced a systemic velocity of $\gamma=-13.9\pm0.8$~\kms\ and a velocity amplitude of $K_1=50.0\pm2.1$~\kms\ for the primary and $K_2=78.1\pm5.7$~\kms\ for the secondary.  This system has a projected semi-major axis of $a\sin i=0.07$~AU.  \cite{Hambleton2013} note that the light curve shows an apsidal advance over the four years of \emph{Kepler} observations, which may lead to an understanding of the binary evolutionary state. 

Line splitting in this system occurs only near periastron, so just two measurements of the secondary velocity can be obtained with our spectra.  These data are enough, though, to place good constraints on the mass of the secondary star.  The mass ratio is measured to be $q=0.64\pm0.09$ and $(M_1+M_2)\sin^3 i$, the sum of the masses times the sine of the inclination cubed, is $1.9\pm0.3$. If we adopt an F0V spectral type corresponding to a mass of 1.6~\msun\ for the primary, we find that the secondary star has a mass of 1.0~\msun.  This corresponds roughly to an G1V spectral type. The inclination of the system can then be calculated to lie in the range 51--65\deg, significantly higher than TH12's reported inclination of 38\deg.

If the primary is an F0V star with a G1V secondary, we would expect roughly 80\% of the flux to come from the primary and 20\% from the secondary.  The redshifted component we see at periastron has an equivalent width in \ion{Si}{2} $\lambda$6371 that is about 20--25\% of the total equivalent width we see in a blended spectrum.  Thus, the secondary mass we derive above is consistent with the flux ratios seen in our spectra.

\subsection{\K1}\label{k1}

\begin{figure}
\centering
\epsscale{1.0}
\plotone{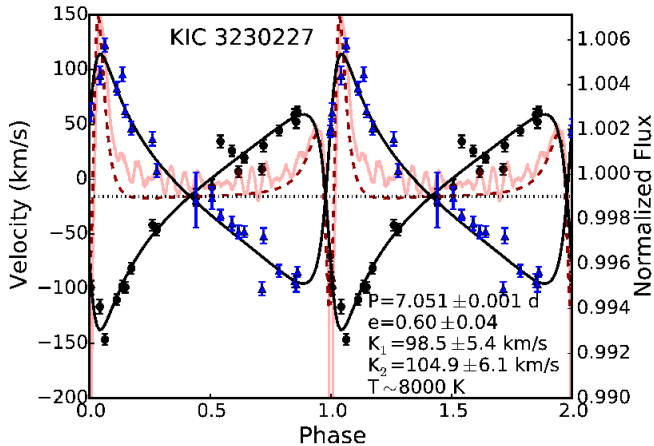}
\caption{The radial velocity data and best-fit solution for \K1, a 7.051 day period SB2 system.  Circles denote the primary, while triangles are the data for the secondary component.  This figure follows the same conventions as Figure~\ref{K424soln}. The grazing eclipse at phase=1.0 (which has a 6\% depth in the full light curve) has been truncated 1\% to allow the stellar pulsations to be seen. (A color version of this figure is available in the online journal.)
\label{K323soln}} 
\end{figure}

Our spectra reveal that \K1 is a double-lined spectroscopic binary.  TH12 reported grazing eclipses in the light curve but did not note the double-lined behavior.  Our observations include 23 nights from 2012 March through 2013 June. We find that the two components are of roughly equal temperature, estimated to be 8000~K (discussed below), while TH12 estimated a temperature for the system closer to 8750~K. 

For \K1\ our best-fit solution provides a period of $P=7.051\pm0.001$~days.  This agrees well with the TH12 value of 7.047~days.  We find an eccentricity of $e=0.60\pm0.04$, in good agreement with TH12's value of 0.59.  The velocity amplitude of the primary is $K_1=98.5\pm5.4$~\kms, while the secondary has an amplitude of $K_2=104.9\pm6.1$~\kms. The similarity of the velocity amplitudes indicates a mass ratio very close to unity. We find a systemic velocity of $\gamma=-15.7\pm1.8$~\kms\ and a projected semi-major axis of $a\sin i=0.05$~AU. Figure~\ref{K323soln} shows the best-fit solution and measured velocities.  Circles denote data from the primary while triangles denote data from the secondary.  There is a 6\% decrease in normalized flux at the primary eclipse; the secondary eclipse is not seen owing to the highly eccentric orbit.  The primary eclipse has been truncated at 1\% of its observed depth in Figure~\ref{K323soln} to allow the stellar pulsations to be more easily seen.  Our analytic light curve model (dashed line) does not include an eclipse and may have a small scaling error due to uncertainties in the depth of the tidal variation.  

\begin{figure}
\epsscale{1.0}
\centering
\plotone{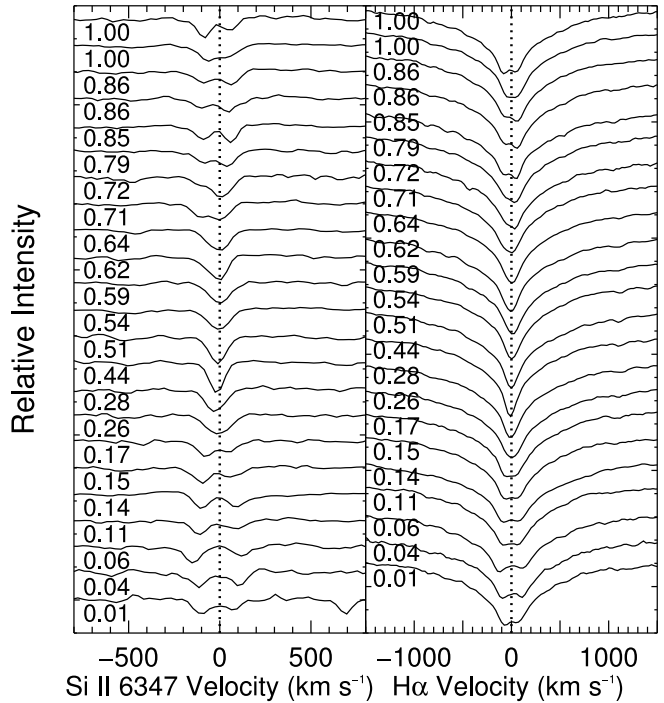}
\caption{Profiles of the \ion{Si}{2} $\lambda$6347 feature and H$\alpha$ in \K1 when ordered by phase (labeled under each spectrum).  The horizontal axis shows velocity centered at the rest wavelength of the feature, as denoted by the vertical dotted line. Both features show obvious splitting at phases near periastron passage (0 and 1). 
\label{lines}} 
\end{figure}

Figure~\ref{lines} shows a time series of the \ion{Si}{2} $\lambda$6347 and H$\alpha$ lines  ordered by orbital phase with the horizontal axis in \kms\ and the dotted line at 0~\kms\ showing the rest wavelength.  Both lines display an obvious splitting near phase=0.  This splitting is mimicked in other lines, such as the \ion{Si}{2} $\lambda$6371 feature.  Equivalent widths of the silicon and H$\alpha$ features, as well as other lines not pictured, appear slightly smaller in the secondary component, but the spectral types must be similar because the spectral features are nearly identical in unblended spectra.

\subsubsection{Temperature}\label{k1t}

Because of the difficulty in calculating temperatures for the individual components of \K1, we decided to combine model atmospheres in an attempt to reproduce the observed spectrum at periastron when the two stellar components are minimally blended.  We used the PHOENIX model atmospheres of \cite{Husser} because the spectral resolution is very similar to that of our data and the temperature grid spans the range of A-type stars  in increments of 200~K.  We corrected the model spectra to air wavelengths and continuum normalized the flux in a manner similar to observed spectra.  We shifted the blue component to $-$146.5~\kms\ and the red component to 122.6~\kms\ to reflect the maximum velocities in our observed spectra.  We also added in an empirical \ion{Na}{1}~D profile modeled after the interstellar sodium absorption of a spectrum where the lines are fully blended.  Finally, we multiplied our resultant spectrum by the atmospheric transmission spectrum from \cite{Hinkle}. We find the atmospheric transmission necessary to include because the H$_2$O feature near the sodium doublet and the O$_2$ feature close to the silicon lines, to name two, affect the continuum and depth of the observed features.  Although the atmospheric transmission at Kitt Peak measured by \citet{Hinkle} may not exactly match that at WIRO, this zeroth order correction is sufficient to show the probable magnitude of the atmospheric features in the spectra.

\begin{figure*}
\centering
\plotone{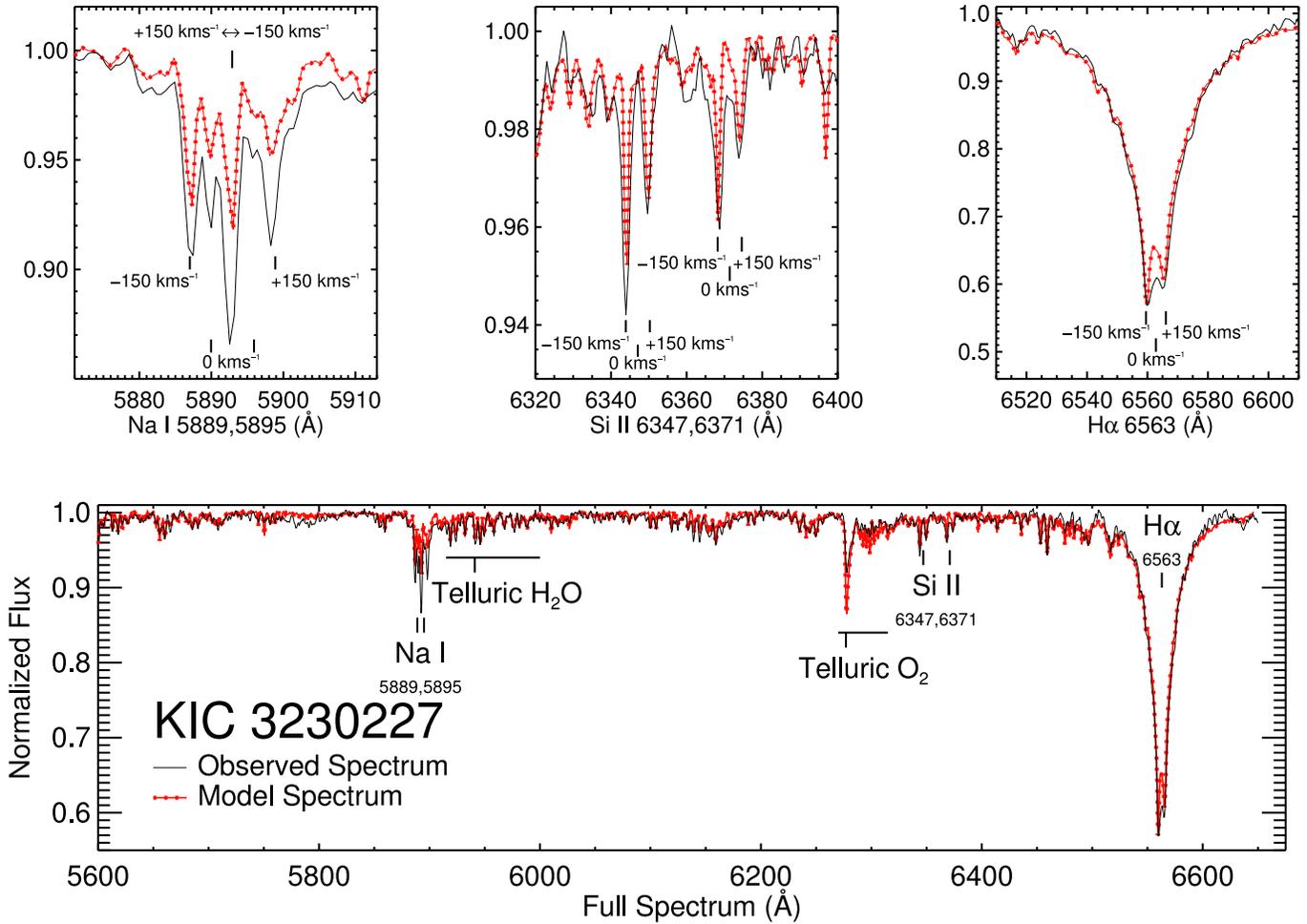}
\caption{The best-fit spectrum for \K1.  The bold line depicts the composite model spectrum, consisting of a blueshifted model stellar atmosphere with a temperature of 8000~K and a surface gravity of $\log g=3.5$, a redshifted model atmosphere with T=8000~K and $\log g=4.0$, an empirical model of the interstellar sodium doublet \ion{Na}{1}~D $\lambda\lambda$5889,5995, and the atmospheric transmission spectrum of \cite{Hinkle} smoothed to the resolution of the data.  The thin black line shows a spectrum taken with WIRO when the components are at maximum separation.  The top three panes show magnified versions of the \ion{Na}{1}~D doublet, the \ion{Si}{2} lines, and H$\alpha$, respectively.  In these top panels, the rest wavelength of each feature has been marked ``0~\kms''.  The wavelengths at $\pm150$~\kms\ have also been identified. Note that, in the \ion{Na}{1}~D panel, the redshifted component of the $\lambda$5889 feature and the blueshifted component of the $\lambda$5895 feature overlap.  The bottom panel shows the full spectrum with the above three features marked.  We also label two atmospheric regions (H$_2$O and O$_2$) where the observed spectrum is noticeably affected by uncorrected telluric absorption.  (A color version of this figure is available in the online journal.)
\label{K323Tfit}} 
\end{figure*}

Figure~\ref{K323Tfit} shows the best-fit synthetic model spectrum (bold and dotted line; colored red in the electronic edition) overplotted on the observed spectrum (thin black line).  The lower panel shows the full spectral range of our data; the upper three panels show magnified views of important features: \ion{Na}{1} on the left, \ion{Si}{2} in the middle, and H$\alpha$ on the right.  The rest wavelength of each line has been marked as ``0~\kms'', and the velocities of $\pm$150~\kms\ are also labeled.  In the \ion{Na}{1} panel, the redshifted component of the $\lambda$5889 line and the blueshifted component of the $\lambda$5895 line overlap. From our best-fit synthetic spectrum, we estimate that the stars have very similar spectral types near  A6--7V.  Both model atmospheres shown have a temperature of 8000~K.  The redshifted secondary component has a surface gravity of $\log g=3.5$ and rotational broadening of 75~\kms, while the blueshifted primary component has a surface gravity of $\log g=4.0$ and rotational broadening of 30~\kms.   Overall, the figure shows good agreement between the synthetic spectrum and the observed spectrum, especially in regions like H$\alpha$ and \ion{Si}{2}.  However, although the \emph{ratios} of the sodium lines in the model atmosphere match those in the observed spectrum, we have not been able to replicate the \emph{absolute} depth despite having varied parameters such as rotational broadening, spectral type, and surface gravity. Increasing the contribution of the interstellar Na does not improve the fit because the interstellar lines fall at zero velocity and are much weaker than the stellar lines.  Adopting the hotter effective temperature found by TH12 only serves to exacerbate the discrepancy between the synthetic and observed spectra. Increasing the metallicity to [+0.5] and [+1.0] resulted in a slightly better fit to the sodium lines but a much poorer fit to the silicon and other metal lines, in the sense that the metal-rich models greatly overpredicted the strength of metal lines throughout the spectrum. Thus, it is possible that these stars are more metal rich than solar or have peculiar metal abundances, but the difference is not reproducible with the available PHOENIX spectra.

\subsubsection{Inclination}\label{k1i}

Knowing the orbital parameters for the stars of \K1, we calculate that the sum of the masses times $\sin^3 i$ is $3.1\pm0.3$~\msun\ and that the mass ratio is $q=M_2/M_1=0.94\pm 0.05$. Adopting TH12's inclination of 43\deg, the sum of the masses would be $\sim$10~\msun.  This would correspond to two late-type B stars, in strong disagreement with the fitted temperatures and spectral types.  An inclination of 66--71\deg\ is required to yield a sum of $\sim$4~\msun\ appropriate to two mid-type A stars.  Inclinations as low as 55\deg\ and as high as 80\deg\ may be allowed owing to the uncertainties on the components' spectral types.  TH12 note that their calculated inclination for \K1 seems low for an eclipsing system and that it may be underestimated.  The most likely reason, according to their analysis, is because the model from \cite{Kumar1995} does not include stellar irradiation.

\section{Discussion and Summary}\label{conclusion}

We have presented spectroscopic orbital solutions for six \hb, further supporting the hypotheses of \cite{Welsh2011}, \cite{Thompson2012}, and others that this class of stars with highly variable light curves are indeed high-eccentricity binary systems.  All six systems have best-fit eccentricities $>0.34$ (most above $\sim0.5$) and periods of 7--20 days. The good agreement between our spectroscopic orbital solutions and the solutions derived from \emph{Kepler} light curves in TH12 attests to the applicability of the \cite{Kumar1995} model of static modes of gravitationally-induced tidal variations to \hb\ orbital element extraction.

Each system in our sample contains an A- or F-type primary component.  The six systems, of which two are SB2s and four are SB1s, have minimum mass ratios $q \approx 0.3$--1.0, making the secondaries probable M-dwarf stars or earlier. \K1 is a double-lined system containing two mid-A stars with a mass ratio very close to unity.  This system is reminiscent of KOI-54, the prototypical heartbeat system \citep{Welsh2011}, which consists of two massive A-type stars in a 41-day orbit. Our other SB2, \K2, exhibits similarities to \K1, such as an eccentricity on the higher end of our sample, but has a smaller primary mass, longer period, and smaller mass ratio.  The two SB2s have the highest mass ratios of our sample, and both SB2s have inclinations higher than reported in TH12. This may be an indication of stellar irradiation or other effects creating signatures in the light curve that are not included in the \cite{Kumar1995} model.  These limited data suggest that photometrically derived orbital elements systematically underestimate the inclination for high mass ratio systems, although more data would be required to confirm this observation. The \hb\ herein exhibit mass ratios $q\gtrsim 0.3$, suggesting that relatively massive companions may be required to excite the tidal and pulsational modes that are the hallmark of these intriguing systems.  However, a wide range of companions are allowed.  On the basis of this limited sample size, there does not appear to be evidence for a correlation between primary mass and mass ratio or between orbital elements and mass ratio.  These spectroscopic observations provide fundamental data on the masses and mass ratios of \hb\ that can be used to further understand their complex photometric variability, orbital dynamics, and binary evolution. 

\acknowledgments  We thank our anonymous referee for comments that greatly improved this work. We thank Martin Still for a timely conversation at the 2012 winter AAS meeting that inspired this project and for encouraging comments on an early version of this manuscript. We thank Daniel Kiminki for his assistance with software and analysis of \K1.  For their observing help, we also thank Michael J. Lundquist, Jamison Burke, James Chapman, Erica Keller, Kathryn Lester, Emily K. Rolen, Eric Topel, Arika Egan, Alan Hatlestad, Laura Herzog, Andrew Leung, Jacob McLane, Christopher Phenicie, and Jareth Roberts. We acknowledge continued support from the Wyoming NASA Space Grant Consortium through NASA Grant \#NNX10A095H.  We are grateful to WIRO staff James Weger and Jerry Bucher whose diligent support work enabled this program to obtain spectra  at the Wyoming Infrared Observatory 2.3 m telescope. Some of the data presented in this paper were obtained from the Mikulski Archive for Space Telescopes (MAST). STScI is operated by the Association of Universities for Research in Astronomy, Inc., under NASA contract NAS5-26555. Support for MAST for non-HST data is provided by the NASA Office of Space Science via grant NNX13AC07G and by other grants and contracts.

\textit{Facilities:} \facility{WIRO ()}

\end{document}